\documentclass[11pt, oneside]{article}   	
\usepackage[margin=1in]{geometry}               
\usepackage{graphicx}				
										
\usepackage{amssymb}
\usepackage{physics}
\usepackage{amsmath}
\usepackage[dvips]{epsfig}

\usepackage{amsfonts}
\usepackage{subfig}

\usepackage{mathrsfs}

\usepackage{xcolor}

\usepackage{bm}

\usepackage{blindtext}

\usepackage[title]{appendix}

\usepackage{cite}

\usepackage{authblk}

\def\cchi{\raise2pt\hbox{$\chi$}} %slightly higher chi

\title{\bf{Quantum Lattice Representation of Nonlinear Classical Physics}}

\author {Min Soe ${ }^{1}$, George Vahala${ }^{2}$, Linda Vahala ${ }^{3}$, Abhay K. Ram ${ }^{4}$, Efstratios Koukoutsis ${ }^{5}$, Kyriakos Hizanidis ${ }^{5}$\\
${ }^{1}$ Department of Mathematics and Physical Sciences, Rogers State University, Claremore,OK 74017\\
${ }^{2}$ Department of Physics, William \& Mary, Williamsburg, VA23185\\
${ }^{3}$ Department of Electrical \& Computer Engineering, Old Dominion University, Norfolk, VA 23529\\
${ }^{4}$ Plasma Science and Fusion Center, MIT, Cambridge, MA 02139\\
${ }^{5}$ School of Electrical and Computer Engineering, National Technical University of Athens,Zographou 15780, Greece \\}
\date{}

\begin{document}
\maketitle
%\section{}
\begin{abstract}
Using the Madelung transformation on a generalized scalar Gross-Pitaevski equation, a nonlinear continuum fluid equations are derived for a classical fluid.  A unitary quantum lattice algorithm is then determined as a second order discrete representation of this Gross-Pitaevski equation and the simulations are compared to those using classical fluid dynamic techniques.
%A Dyson map explicitly determines the appropriate basis of electromagnetic fields which yields a unitary representation of the Maxwell equations in
%an inhomogeneous medium.
%A qubit lattice algorithm (QLA) is then developed perturbatively to solve this representation of Maxwell equations.  A
%QLA consists of an interleaved unitary sequence of collision operators (that entangle on lattice-site qubits) and streaming operators (that move this entanglement throughout the lattice).
 %External potential operators are introduced to handle gradients in the refractive indices, and these operators are typically non-unitary but sparse matrices.  By also interleaving the external potential operators with the unitary collide-stream operators, one achieves a QLA which conserves energy to high accuracy.  
 %Some two dimensional simulations results are presented for the scattering of a one-dimensional (1D) pulse off a 
 %localized anisotropic dielectric object.

%  data :  /Users/jiri/2023.JANUARY.ANISO.CYL/POLARIZATIONS

\end{abstract}

\section{Quantum Information Science and Nonlinear Classical Physics}

  $\quad$ In theory, it is expected that quantum computers can provide exponential speedup over classical supercomputers for certain physics problems.
This leads to the question of how quantum information science (QIS) can play a role in solving nonlinear problems, since quantum mechanics is a linear theory.
In particular, the nonlinear equations of computational fluid dynamics (CFD) pose a serious challenge in the development of an QIS algorithm [1-9].  
The specific QIS hurdles involve not only how to handle nonlinear effects, but also the question of obtaining time evolution data from a quantum computer
that will satisfy CFD solutions.  A quantum measurement not only will require multiple realizations of the quantum algorithm, but the effect of the measurement
will destroy the measured quantum state of the computation at that instant so that time evolution classical data will require reinitialization of the quantum bits
following every measurement.  This is very destructive on the quantum speed-up that can be achieved in a QIS representation.
Handling the nonlinear CFD terms like the convective derivative in the Navier-Stokes equation $\mathbf{u} \cdot \nabla \mathbf{u}$, where $\mathbf{u}$ is
the classical fluid velocity, in QIS hits against the no-cloning theory [3]:  in QIS one cannot make an exact copy of an unknown quantum state.
Any measurement of that quantum state will destroy all quantum parallelization advantages of QIS over CFD and then require reinitialization of the qubit states
to continue the QIS simulation.

Carleman [4] introduced the idea of transforming a nonlinear equation into a set of linear coupled equations.   However, there is a closure problem:  one has to deal with an infinite set of coupled linear equations.  For fully nonlinear problems like fluid turbulence, a truncated finite set of coupled equations does not lead
to a convergent representation of the solution.  In particular, the QIS Carleman representation of the Burgers equation falls apart as the nonlinearity tends to
a steepening shock [5].  Attempts have been made recently by Succi and his co-workers on applying the Carleman formulation to the quantum lattice
Boltzmann representation of classical fluid turbulence.  The idea was that the convective derivative in Navier-Stokes turbulence is modeled by simple advection 
in lattice Boltzmann with the nonlinearity showing up as a quadratic term in the relaxation distribution function.  The Carleman linearization may thus be more
potent on the quantum lattice Boltzmann representation or on the Grad 13-moment model [6].

Other quantum attempts at handling the nonlinear equations of fluid mechanics involves a hybrid quantum-classical approach using variational methods [7], or evaluating the nonlinear terms on a computational basis by converting the solution determined in the amplitude representation using an analog-to-digital 
routine.  Classical techniques are then used to calculate the nonlinear terms which are then transformed by digital-to-analog routine into a quantum amplitude basis [8].  Here we will consider a somewhat different approach, using the interplay between the Gross-Pitaevskii equation for the nonlinear evolution of the quantum ground state of a Bose-Einstein condensate (BEC) and the corresponding fluid equations determined by the Madelung transformation,
 as pointed out by Nore et. al. [9].   Indeed, the scalar 3D GP equation
(with an effective potential V)
\begin{equation}
i \frac{\partial \psi}{\partial t} = \left( -\nabla^2  +V  + g |\psi|^2 \right)  \psi
\end{equation}
under the Madelung transformation
\begin{equation}
\psi = \sqrt{\rho} \,e^{i \theta /2}
\end{equation}
yield the conservation equations of mass and momentum% where the fluid density $\rho = |\psi|^2$, and fluid velocity 
%$\mathbf{u}=\nabla{\theta}$:
\begin{equation}
\frac{\partial \rho}{\partial t} + \partial_j \left({\rho u_j} \right ) = 0 ,
\end{equation}

\begin{equation}
\frac{\partial \rho u_i}{\partial t} + \partial_j \left [\rho u_i u_j + p \delta_{i,j} \right]  = - \partial_i \left[ 4 \, \partial_j \sqrt{\rho} \partial_i \sqrt{\rho} - \nabla^2 \rho \, \delta_{i_j}  \right] ,
\end{equation}
where the fluid density $\rho = |\psi|^2$, fluid velocity 
$\mathbf{u}=\nabla{\theta}$, and p is the fluid pressure.  We do not need to consider explicitly the conservation of fluid energy.  The left hand side of the momentum equation (4) is that for a classical barotropic fluid, while the gradients of the density on the right hand side of (4) are the quantum effects  (leading to a quantum pressure).
Since the GP equation is conservative, no transport (dissipative) coefficients appear in the fluid momentum equation (4).

A scalar quantum vortex is a singular solution to the GP equation (1), with $\rho \rightarrow 0$ at the vortex core.  Thus a quantum vortex requires
a compressible fluid.   A unitary QIS representation of the GP equation (1) has been determined using the qubit lattice algorithm (QLA) [2].  In QLA one defines a sequence of non-commuting unitary collision and streaming operators that in the continuum limit recovers the GP Eq. (1) to second order
in the lattice grid spacing.  One qubit/lattice site is introduced for the scalar wave function $\psi$, so that the collision and streaming unitary  operators
are $2 \cross 2$ matrices.  In CFD turbulence studies, one of the major hurdles to be cleared is the accurate discrete representation of the nonlinear
convective derivative terms $\partial_j \left [\rho u_i u_j + p \delta_{i,j} \right] $.  Here we see that QLA has transformed the problem into finding the
approrpiate unitary collision and streaming operators [10].

An interesting generalization of the scalar GP equation (1) has been given by Meng and Yang [11].  These authors wanted to determine a QIS algorithm
to solve a particular classical nonlinear fluid system.  In particular they [3] introduced some free functions into their generalized GP Eq. (1) so that under
the Madelung transformation one recovers a purely classical nonlinear fluid set of equations under specific choices of these free functions.  
One immediately sees that all scalar quantum vortices can be eliminated if the free functions can enforce fluid incompressibility, $\rho = const.$   
This is because a scalar quantum vortex requires a compressible core to exist.  Another important consequence of fluid incompressibility is
that the quantum pressure - a term that needs to be excluded in a QIS representation of classical nonlinear fluids - is also eliminated.
Meng and Yang [11] use quaternions in their derivation - but this is exactly what is done in the qubit representation.  After considerable work they can
model the incompressible nonlinear fluid equations
\begin{equation}
\rho = \rho_0 = const.
\end{equation}
\begin{equation}
\frac{\partial \mathbf{u}}{\partial t} + \mathbf{u} \cdot \nabla \mathbf{u} = - \nabla \left(\frac{p}{\rho_0} \right) - \frac{1}{4 \, \rho_0^2} \nabla \mathbf{s} \cdot \nabla^2 \mathbf{s}
\end{equation}
with pressure 
\begin{equation}
p = \frac{1}{4 \, \rho_0} | \nabla \mathbf{s} |^2
\end{equation}
and the spin vector $\mathbf{s}$ which is best defined in terms of the qubit wave function $\boldsymbol{\psi} = [\psi_1 \quad \psi_2]^T$ 
\begin{equation}
\mathbf{s} = \left( |\psi_1|^2 - |\psi_2|^2 \quad , i \left[ \psi_1^\star \psi_2 - c.c \right] \quad, \psi_1^\star \psi_2 + c.c \right) 
\end{equation}
whose time evolution resembles the Landau-Lifshitz-Gilbert equation that occurs in ferromagnets and vortex filaments [4] and which has close 
connections with integrable soliton equations.  One can show that
\begin{equation}
 | \nabla \mathbf{s} |^2 = 4 \rho_0 \left( | \nabla \psi_1|^2  +  | \nabla \psi_2|^2 \right) - \rho_0 | \mathbf{u} |^2
\end{equation}
These nonlinear (conservative) fluid equations are modeled by the generalized GP 2-component wave function
\begin{equation}
i \frac{\partial \boldsymbol{\psi}}{\partial t} = \left( - \nabla^2 + \frac{1}{4 \rho^2} | \nabla \mathbf{s} |^2 \right) \boldsymbol{\psi}
\end{equation}
The generalized spinor GP equation (10) is then solved [11] using CFD methods and results plotted for the evolution of the Taylor-Green vortices in 2D.

\section{QLA representation for the Generalized spinor GP equation (10)}

$\quad$ QLA is a sequence of unitary collide-stream matrices which perturbatively recover the desired nonlinear continuum GP equation.  The perturbative parameter
arises because QLA is a mesoscopic discrete representation based on the spatial lattice grid is of $O(\epsilon)$.
The extension of QLA for the GP Eq. (1) to handle the Meng-Yang generalized GP Eq. (10) is immediate.  Equation (10), like Eq. (1), requires only 1 qubit field to be introduced. The unitary collision operator is
 \begin{equation}
C=\left[\begin{array}{cc}
\cos \gamma &i \sin \gamma \\
i \sin \gamma & \cos \gamma
\end{array}\right]
\end{equation}
where (real) $\gamma$ is defined by
\begin{equation}
\gamma = \frac{\pi}{4} + \epsilon^2 V_{ext} \quad \quad  \text{with} \quad \quad V_{ext} = -\frac{1}{64 \rho_0^2} | \nabla \mathbf{s} |^2.
\end{equation}

The unitary streaming operators in the x-direction: $S_1^{+x}$ streams the $\psi_1$ component one lattice unit to the right, while $S_2^{-x}$ streams the 
$\psi_2$ component to the left one lattice unit.  The 2D unitary evolution operator that will propagate the spinor $\boldsymbol{\psi}(t)$ to 
$\boldsymbol{\psi}(t+\Delta t)$ can be factored into a product of two orthogonal individual unitary evolution operators $\mathbf{U_X}$ and 
$\mathbf{U_Y}$
\begin{equation}
\boldsymbol{\psi(t+\Delta t)} = \mathbf{U_Y \, U_X} \boldsymbol{\psi(t)}
\end{equation}
where
\begin{equation}
\mathbf{U_X} = S_2^{-x} C.S_2^{+x} C.S_2^{-x} C.S_2^{+x} C,\,S_1^{-x} C.S_1^{+x} C.S_1^{-x} C.S_1^{+x} C.
\end{equation}
$\mathbf{U_Y}$ is just Eq. (14), but with  $x \rightarrow y$ in the streaming operators.

The discrete unitary QLA sequence, Eq. (13) and (14), recovers the required continuum GP Eq. (10) under diffusion order ($\Delta t 
=O(\epsilon^2) = \Delta x^2$) to second order in $\epsilon$.

We have performed QLA simulations for the evolution of 2D Taylor-Green vortices with initial mean velocity
\begin{equation}
\mathbf{u} =  \left( sin \,x \, cos \,y \, , -cos \,x \,sin \,y \right)
\end{equation}
and find very good agreement with the vorticity fields determined  by Meng-Yang [11] using CFD solvers for Eq. (10).  Our QLA simulations 
are presented in Fig. 1

 \begin{figure}[!h!p!b!t] \ 
\begin{center}
\includegraphics[width=2.1in]{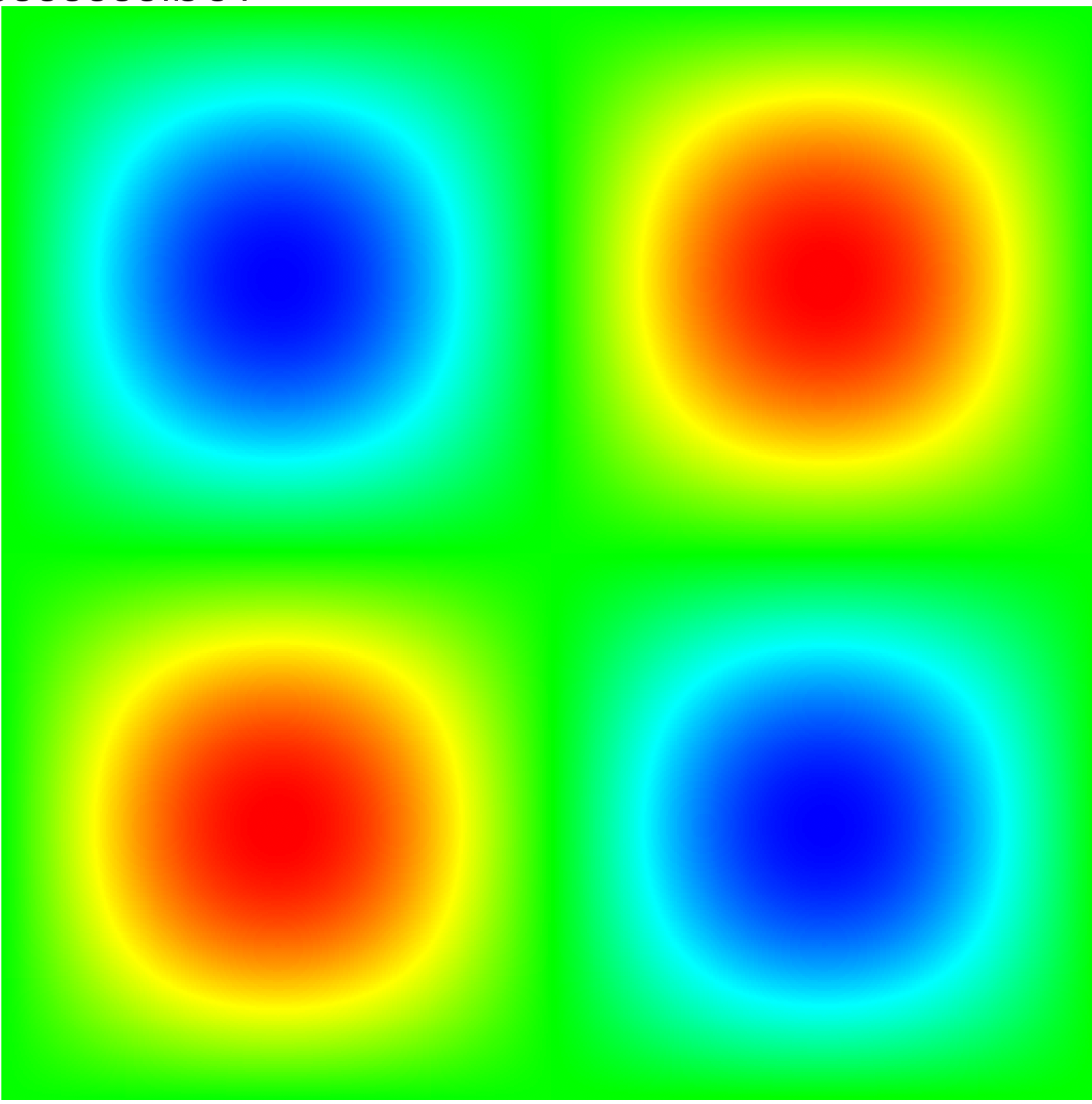}
\includegraphics[width=2.1in]{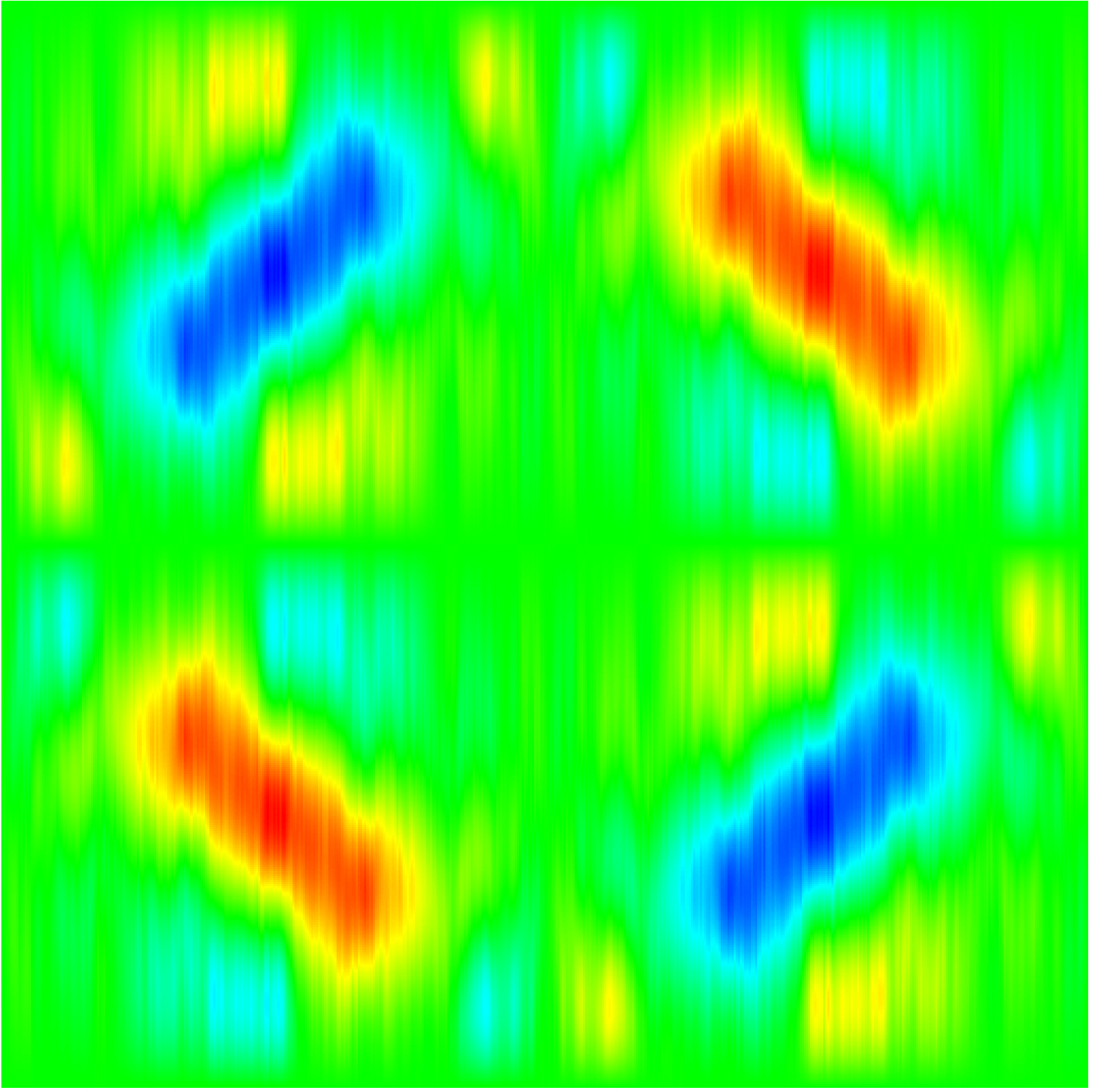}
\includegraphics[width=2.1in]{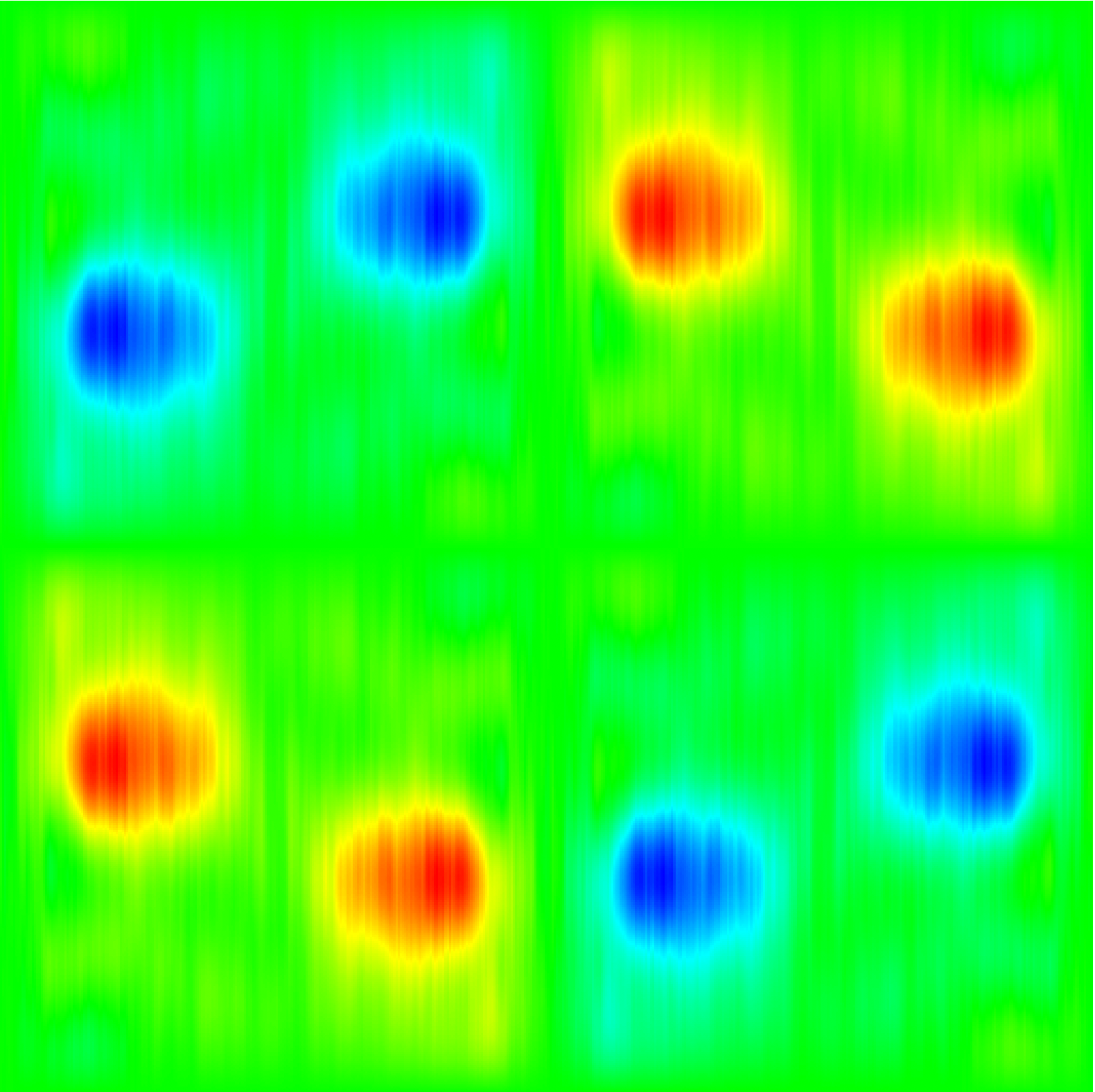}
(a)  time $t_0$   \qquad \qquad  \qquad  (b)  time $ t_1$ \qquad  \qquad  \qquad (c)  time $ t_2$ 
\caption{ The QLA evolution of initial 2D Taylor-Green vortices, (a), following the time evolution of a particular nonlinear conservative fluid
equation, Eq. (10) for successive times (b) and (c).  Red vortices rotate clockwise, while blue vortices rotate anti-clockwise.
}
\end{center}
\end{figure}

 \section{Acknowledgments}
This research was partially supported by Department of Energy grants DE-SC0021647, DE-FG0291ER-54109, DE-SC0021651, DE-SC0021857, and DE-SC0021653. This work has been carried out partially within the framework of the EUROfusion Consortium. E.K has received funding from the Euratom research and training program WPEDU under grant agreement no. 101052200 as well as from the National Program for Controlled Thermonuclear Fusion, Hellenic Republic. K.H is
supported by the National Program for Controlled Thermonuclear Fusion, Hellenic Republic. The views and opinions expressed herein do not necessarily reflect those of the European Commission.   This research used resources of the National Energy Research Scientific Computing Center (NERSC), a U.S. Department of Energy Office of Science User Facility located at Lawrence Berkeley National Laboratory, operated under Contract No. DE-AC02-05CH11231 using NERSC award FES-ERCAP0020430.

\section{References}
[1]  C. Sanavio and S. Succi, "Quantum computing for simulation of fluid dynamics", arXiv:2404.01302v1 (2024).

\noindent[2]  S. Succi, W. Itani, K.Sreenvasan and R. Steijl, "Quantum computing for fluids:  where do we stand?  Europhys. Lett. $\mathbf{144}$, 10001 (2023). 

\noindent[3]  M.A. Nielsen and I. L. Chuang, "Quantum computation and quantum information", 10th Anniversary Edition, Cambridge University Press, (2010).

\noindent[4]  T. Carleman, "Application of the theory of linear integration equations to nonlinear system of differential equations", Acta Mathematica, $\mathbf{59}$, 63 (1932).

\noindent[5]  J. - P. Liu, H. O. Kolden, H. K. Krovi, N. F. Loureiro, K. Trivisa, and A. M. Childs, "Efficient quantum algorithm for dissipative nonlinear 
differential equations", Proc. Natl. Acad. Sci. U. S. A $\mathbf{118}$, e2026805118 (2021)

\noindent[6]  C. Sanavio, S. Succi and E. Mauri, "Carleman-Grad approach to the quantum simulation of fluids", arXiv:2406.01118v1 (2024)

\noindent[7]  M. Lubasch, J. Joo, P. Moinier, M. Kiffner and D. Jaksch, "Variational quantum algorithms for nonlinear problems", Phys. Rev. $\mathbf{A101}$, 010301 (2020).

\noindent[8]  R. Steijl, "Quantum circuit implementation of multi-dimensional non-linear lattice models,"  Appl. Sci.$\mathbf{13}$, 529 (2022).

\noindent[9]  C. Nore, M. Abid, and M. E. Brachet, "Decaying Kolmogorov turbulence in a model of superflow", Phys. Fluids $\mathbf{9}$, 2644-2669 (1997)

\noindent [10]  J. Yepez, G. Vahala, L. Vahala and M. Soe, "Superfluid turbulence from quantum Kelvin wave to classical Kolmogorov cascades". Phys. Rev. Lett. $\mathbf{103}$, 084501 (2009)

\noindent [11]  Z. Meng and Y. Yang, "Quantum computing of fluid dynamics using the hydrodynamic Schrodinger equation", Phys. Rev. Research $\mathbf{5}$, 033182 (2023)

\noindent [12]  M. Lakshmanan, "The fascinating world of the Landau-Lifshitz-Gilbert equation: an overview", Phil Trans Royal Soc. $\mathbf{A369}$, 1280-1300 (2011)

%................

\end{document}